\documentclass[a4paper]{article}
\usepackage{INTERSPEECH2022}
\usepackage{cite}
\usepackage{amsmath,amssymb,amsfonts}
\usepackage{algorithmic}
\usepackage{graphicx}
\usepackage{textcomp}
\usepackage{xcolor}
\usepackage{multirow}
\usepackage{booktabs}
\usepackage{comment}
\usepackage{makecell}
\usepackage{threeparttable}
\usepackage{url}
\usepackage{tablefootnote}

\title{The DKU System for Multi-Speaker Automatic Speech Recognition in MLC-SLM Challenge}
\name{Yuke~Lin$^{1,2}$, Ming~Cheng$^{1,2}$, Ze~Li$^{1,2}$, Ming~Li$^{1,2*}$\thanks{*Corresponding Author}}

\address{
  $^1$School of Computer Science, Wuhan University, China\\
  $^2$Suzhou Municipal Key Laboratory of Multimodal Intelligent Systems, Digital Innovation Research Center, Duke Kunshan University, China}
\email{ming.li369@dukekunshan.edu.cn}

\begin{document}

\maketitle
\begin{abstract}
We present the DKU system for Task 2 of the MLC-SLM Challenge, which aims to perform multi-speaker automatic speech recognition directly from raw audio without Oracle speaker labels or time boundaries. Our approach builds upon a diarization-aware framework integrating speaker embeddings and temporal utterance boundaries into a Qwen2.5-based large language model (LLM). Then, we enhance the system's multilingual performance by fine-tuning language-specific adapters and LoRA modules within the LLM decoder. Finally, our system achieves the tcpWER of 23.56\% and 18.08\% on the development and test sets of the MLC-SLM dataset, substantially outperforming the official baseline.
\end{abstract}

\section{Introduction}

Automatic speech recognition (ASR) has achieved remarkable success in single-speaker scenarios~\cite{asr01, asr02, asr03}, but real-world conversations often involve multiple speakers, overlapping speech, and code-switching, posing significant challenges to current approaches. Multi-speaker ASR (MS-ASR)~\cite{MS-ASR_survey, ms-asr-first-type01, ms-asr-first-type02, ms-asr-second-type01, ms-asr-second-type02, pit01, sot01, t-sot} aims to address the ``who spoke what and when'' problem by jointly recognizing speech content and attributing it to the correct speaker. This task becomes even more difficult in multilingual and conversational settings, where speaker identity, timing, and linguistic diversity must be accurately disentangled.

The Challenge and Workshop on Multilingual Conversational Speech Language Model (MLC-SLM)~\footnote{\url{https://www.nexdata.ai/competition/mlc-slm}} focuses on building robust ASR systems under such realistic conditions. In particular, Task 2 requires systems to jointly perform speaker diarization and transcription from raw, unsegmented audio without access to Oracle speaker labels or time boundaries. The final ranking is based on the time-constrained permutation Word Error Rate (tcpWER)~\cite{tcpwer}, which jointly evaluates recognition accuracy and temporal alignment.

This report describes the DKU team’s system for Task 2 of the MLC-SLM Challenge. The developed system is based on the Diarization-aware Multi-speaker ASR~\cite{lin2025diarization} framework that integrates speaker embeddings and utterance-level timestamps into a large language model~\cite{llama, qwen2.5, deepseek}. Compared to the original paper, we further fine-tune language-specific adapters and LoRA~\cite{lora} modules for each language category to improve the multilingual performance.

\begin{figure}[htbp]
    \centering
    \includegraphics[width=0.9\linewidth]{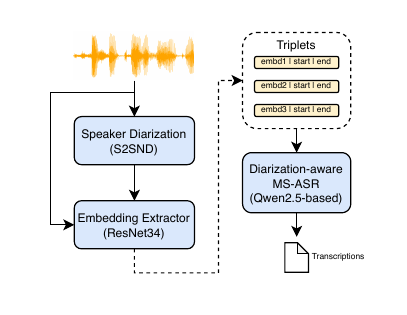}
    \caption{Overview of our developed system.}
    \label{fig:overview}
\end{figure}

\begin{table*}[th]
    \setlength{\tabcolsep}{20pt}
    \renewcommand{\arraystretch}{1.2}
    \centering
    \caption{Performance Comparison (tcpWER \%) on the MLC-SLM Dev and Test sets by different languages. $*$ denote the system fine-tuned by language-specific adapter and LoRA. }
    \label{tab:mlc_slm}
    \begin{tabular}{lcccccc}
    \toprule
    \multirow{2}{*}{\textbf{Language}} 
    & \multicolumn{2}{c}{\textbf{Baseline~\tablefootnote{\url{https://github.com/mubingshen/MLC-SLM-Baseline/tree/main}}}} & \multicolumn{2}{c}{\textbf{MS-ASR~\cite{lin2025diarization}}} & 
    \multicolumn{2}{c}{\textbf{MS-ASR\textsuperscript{*}}} 
    \\
    \cmidrule(lr){2-3}
    \cmidrule(lr){4-5}
     \cmidrule(lr){6-7}
    & \textbf{Dev} 
    & \textbf{Test} 
    & \textbf{Dev} 
    & \textbf{Test}
    & \textbf{Dev} 
    & \textbf{Test}
    \\
    \midrule
    English-American & 53.73 & - & 23.01 & - & 23.52 & - \\
    English-Australian & 52.63 & - & 14.40 & - & 15.37 & - \\
    English-British & 71.92 & - & 18.69 & - & 19.15 & - \\
    English-Filipino   & 50.37 & -  & 18.14 & - & 18.04 & - \\
    English-Indian   & 70.72 & -  & 16.25 & -  & 16.54 & - \\
    French   & 96.04 & -  & 34.74 & - & 29.91 & - \\
    German   & 86.74 & -  & 30.38 & - & 29.92 & - \\
    Italian   & 83.31 & -  & 19.90 & - & 21.10 & - \\
    Japanese   & 71.30 & -  & 36.29 & - & 30.65 & - \\
    Korean   & 59.55 & -  & 27.04 & - & 24.95 & - \\
    Portuguese   & 118.84 & -  & 37.35 & - & 34.77 & - \\
    Russian   & 69.21 & -  & 23.20 & - & 20.65 & - \\
    Spanish   & 75.61 & -  & 23.17 & - & 18.95 & - \\
    Thai   & 83.56 & -  & 20.93 & - & 20.93 & - \\
    Vietnamese   & 82.80 & -  & 29.76 & - & 27.75 & - \\
    \midrule
    Overall & 76.12 & 60.39 & 24.95 & 20.44 & 23.56 & 18.08 \\
    \bottomrule
\end{tabular}
\end{table*}

\section{System Description}

\subsection{Overall Architecture}

Our system adopts a diarization-aware pipeline architecture for multi-speaker automatic speech recognition, shown in Fig.~\ref{fig:overview}. The system comprises three main components: the speaker diarization model, the embedding extractor, and the diarization-aware multi-speaker ASR system.

The input audio is firstly processed by a Sequence-to-Sequence Neural Diarization (S2SND) model~\cite{s2snd} to identify utterance-level speaker time boundaries. These time intervals extract speaker-specific audio segments, from which a pre-trained ResNet34~\cite{resnet34} speaker verification model computes utterance-level speaker embeddings. The combination of speaker embedding, start time, and end time forms a set of triplets representing ``who spoke when." 

These triplets are then fed into the MS-ASR module based on the Qwen2.5 language model~\cite{qwen2.5}. This module consists of a semantic encoder (Whisper-large-v3)~\cite{whisper-large-v3}, a speaker feature encoder (the same ResNet34), and a gated attention fusion layer that prepares inputs for the LLM decoder. The model takes the raw audio and multiple triplets as input, which is trained to generate the transcription corresponding to each speaker’s utterance within the specified time window. Each triplet serves as a reference instruction that guides the model to focus on a particular speaker in a defined period.

\subsection{Speaker Diarization}

We adopt the S2SND-Small variant (16.56M parameters)~\cite{s2snd} to perform speaker diarization. It supports online/offline inference and directly predicts utterance-level time boundaries without requiring Oracle VAD or speaker labels. For this challenge, the model is trained on simulated multi-speaker mixtures by the VoxCeleb2~\cite{voxceleb2} dataset and further fine-tuned on the MLC-SLM dataset. Finally, the diarization system achieves an offline DER of 14.27\% on the MLC-SLM development set, which is used in the following embedding extraction and triplet construction.

\subsection{Embedding Extractor}

Speaker embeddings are extracted using a ResNet34~\cite{resnet34} model pre-trained on the VoxBlink2~\cite{voxblink2} dataset for speaker verification. The model is kept frozen during the pipeline and used solely for inference. It achieves an equal error rate (EER) of 1.52\% on the Vox-O trial. For each diarized segment produced by the S2SND module, the corresponding audio span is cropped and passed through the ResNet34 network to obtain an utterance-level speaker embedding. Each embedding is combined with its associated start and end time to form a speaker-enrollment triplet, which serves as a reference input to guide the downstream transcription module.

\subsection{Diarization-aware MS-ASR}

The Diarization-aware Multi-speaker ASR (MS-ASR) module takes two inputs: the full audio waveform and a set of speaker-enrollment triplets. The audio is encoded in parallel by two frozen encoders: a Whisper-large-v3 model~\cite{whisper-large-v3} for extracting frame-level semantic features and a ResNet34 model for extracting frame-level speaker-discriminative features. These two feature streams are fused via a gated cross-attention mechanism, and the result is projected through an adapter before being passed into a Qwen2.5-based language model~\cite{qwen2.5}.

Then, each speaker-enrollment triplet consists of a speaker embedding and a corresponding time interval, used as a reference instruction. During training, the model learns to transcribe the speech content spoken by each specified speaker within the indicated period. Multiple triplets can be processed jointly in a single decoding pass. Importantly, only the adapters and LoRA modules for the LLM are fine-tuned; all other components, including the LLM backbone, Whisper encoder, and ResNet encoder, remain frozen. 

The implementation details, as well as the training and inference processes of the MS-ASR model, are entirely consistent with its original paper~\cite{lin2025diarization}. To further enhance multilingual performance in the context of the MLC-SLM Challenge, we additionally fine-tune separate adapter and LoRA parameters for each target language. Since the language identity is provided as part of the task specification, the model can be controlled to select the appropriate adapter and LoRA branch based on the input language category. Compared to the original system, this extension enables more effective specialization for each language to improve the multilingual recognition performance.

\section{Experimental Results}

Table~\ref{tab:mlc_slm} presents the tcpWER results of three systems evaluated on the MLC-SLM dataset: the official baseline, the original version of MS-ASR model~\cite{lin2025diarization}, and the language-specific MS-ASR\textsuperscript{*} model developed for this challenge. All systems are evaluated on the development set, while only average scores are available on the test set via the official evaluation server.

Compared to the baseline system, which yields an average tcpWER of 76.12\% on the development set and 60.39\% on the test set, the original MS-ASR model achieves substantial improvements, reducing the error to 24.95\% and 20.44\%, respectively. This confirms the effectiveness of the Diarization-aware MS-ASR framework based on the triplet decoding strategy. Further improvements are observed with the language-specific MS-ASR\textsuperscript{*} model, which achieves 23.56\% on the development set and 18.08\% on the test set. The gains are especially notable in languages such as French, Japanese, Russian, and Spanish. However, in high-resource languages like English or low-resource languages already performing well in the base model (e.g., Thai, Italian), the improvements are marginal or slightly regressive, suggesting a possible saturation effect or domain mismatch in the fine-tuning data.

Overall, the results demonstrate that while the basic MS-ASR model provides a strong general solution, incorporating language-specific adaptation can lead to further improvements generally.

\section{Conclusions}

In this work, we build a ASR system for the MLC-SLM Challenge based on the Diarization-aware Multi-speaker ASR framework. The system first performs speaker diarization and embedding extraction to obtain speaker-enrollment triplets, which are then used to guide a large language model for speaker-attributed transcription. For this challenge, we additionally fine-tune language-specific adapters and LoRA modules of the language model to improve its multilingual performance. The final system achieves strong tcpWER results on both the development and test sets.

\bibliographystyle{IEEEtran}
\bibliography{refs}

\end{document}